\documentclass[a4paper,10pt]{article}

\usepackage{graphicx}
\usepackage{natbib}

\usepackage{amsmath}
\usepackage{amssymb}
\usepackage[usenames,dvipsnames]{color} 
\usepackage{bm}
\usepackage{ar} 
\usepackage[utf8]{inputenc}
\usepackage[T1]{fontenc}
\usepackage{textcomp}
\usepackage{gensymb}
\usepackage{authblk}

\newsavebox{\astrutbox}
\sbox{\astrutbox}{\rule[-5pt]{0pt}{20pt}}

\title{Effect of slip on circulation inside a droplet}

\author[1]{Joseph J. Thalakkottor}             
\author[1,2]{K. Mohseni\thanks{Email address for correspondence: mohseni@ufl.edu}}

\affil[1]{Department of Mechanical and Aerospace Engineering, University of Florida, Gainesville, Florida 32611, USA}
\affil[2]{Department of Electrical and Computer Engineering, University of Florida, Gainesville, Florida 32611, USA}

\begin{document}

\maketitle

\begin{abstract}
Internal recirculation in a moving droplet plays an important role in several droplet-based microfluidic devices as it enhances mixing, chemical reaction and heat transfer. The occurrence of fluid slip at the wall, which becomes prominent at high shear rates and lower length scales, results in a significant change in droplet circulation. Using molecular dynamics (MD) simulations, the presence of circulation in droplets is demonstrated and quantified. Circulation is shown to vary inversely with slip length, which is a measure of interface wettability. A simple circulation model is established that captures the effect of slip on droplet circulation. Scaling parameters for circulation and slip length are identified from the circulation model which leads to the collapse of data for droplets with varying aspect ratio (\AR) and slip length. The model is validated using continuum and MD simulations and is shown to be accurate for droplets with high \AR.
\end{abstract}

\section{Introduction}
Circulation in a droplet is known to play an important role in several droplet-based applications due to the enhanced mixing, chemical reaction and heat transfer that results from it. It has an increased impact at small scales where surface effects start to dominate, as seen in several  microfluidic applications such as lab-on-a-chip, microreactors and digitised heat transfer. In a microchannel, mixing or chemical reaction occurs mostly by diffusion. Since the flow remains laminar due to viscous forces dominating over body or inertia forces, fluid circulation or vortices could enhance mixing and/or chemical reactions. This helps in improving the efficiency of miniature-sized bio/chemical analysis systems that use microfluidic devices, such as lab-on-a-chip or micro total analysis systems ($\mu$TAS) \citep{AndreasM:02a,AndreasM:02b}. Circulation also increases  mass transfer, which enhances diffusive penetration and consequently increases the 
observed reaction rates in microreactors. Microreactor technology offers numerous potential benefits for the process industries \citep{RamshawC:01a,MorgenschweisK:03a}. In the case of digitised heat transfer, discrete microdroplets are used to ``digitally'' transfer heat away from the source \citep{Mohseni:05u,Mohseni:08g}. Presence of a fluid-fluid interface leads to the formation of a circulatory flow inside the droplet. This results in convection normal to the wall, improving the efficiency of the thermal management system.  Some research groups have already demonstrated rapid mixing or reaction by means of shuttling a droplet in a microchannel \citep{IsaoE:99a} or transporting it through a winding microchannel \citep{IsmagilovRF:03a}. Thereby, the importance of circulatory flow in microfluidic devices is observed.

Although, slip at the boundary is prevalent for single phase flows, it is negligibly small in most continuum and macro-scale applications. However, in many micro- and/or nano-scale applications the first breakdown of continuum assumption often occurs at a solid boundary in the form of velocity slip. Slip at micro scale has been investigated extensively for single phase fluids, but the research has been limited for effects of slip in two phase flows. Recently, \citet{Mohseni:13e} extended Maxwell's slip model from steady to unsteady flows. The established unsteady model for single phase flows showed a dependence on gradient of shear rate in addition to shear rate of the flow. Also, a time-dependent universal curve for slip at the wall that encapsulated unsteady and steady flows was found. 

The presence of circulatory flow inside a droplet is known for continuum scale problems \citep{PamulaVK:03b,Mohseni:07h}, but unlike some continuum scale hydrodynamic phenomena, circulation is observed even at nanoscales. \citet{KoplikJ:89a} and \citet{ThompsonP:89a} studied the motion of a  contact line using molecular dynamics (MD) simulations and observed the presence of circulatory flow. Here, these observations are replicated and also quantified. Changes in circulation is studied for varying degree of wetting of the wall. It is seen that the occurrence of fluid slip at the wall results in a significant change in circulation. Circulation varies inversely with slip length which could be considered as a measure of interface wettability. In this paper a simple circulation model is established that captures the effect of slip on circulation. Scaling circulation and slip length, the results for different droplet aspect ratio $(\AR)$ collapse onto a single 
curve. The model is verified using results from continuum and MD simulations and is shown to be accurate for droplets with high \AR. It is shown that there can be as much as $50\%$ reduction in circulation for a droplet moving in a non-wetting channel. Finally, the effect of slip at the fluid-fluid interface on circulation in a droplet is briefly discussed.

Details of the problem setup are specified in section~\ref{sec:prob_setup}. In section~\ref{sec:circ_model} the circulation model is derived and the results are discussed in section~\ref{sec:results}.

\section{Problem setup}\label{sec:prob_setup}
\begin{figure}
\centerline{\includegraphics[width=0.43\textwidth]{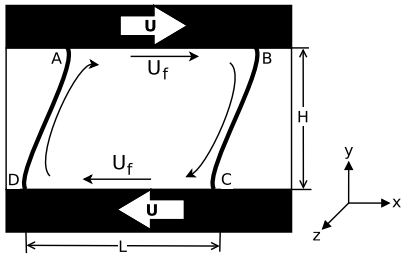}}
\caption{Schematic of the problem. The problem simulates a Couette flow with two immiscible fluids.}
\label{fig:Schematic} 
\end{figure}

In order to study the effects of wall slip on circulation, a droplet moving in contact with a wall is simulated using Couette flow as a canonical problem. Two immiscible fluids occupying equal volume are placed in between two parallel walls separated by a distance $H$. Top and bottom walls move in opposite directions with a velocity $U$. A schematic of the problem geometry is shown in figure~\ref{fig:Schematic}. Periodic boundary conditions are imposed along the $x$ and $z$ directions. The size of fluid domain used in MD simulations, for different droplet aspect ratios are listed in table~\ref{tab:ar_h}. The problem is studied primarily using MD simulations. Due to the high cost of MD simulations, results from a Navier Stokes continuum solver are used to supplement the data, by performing runs for a wider range of droplet aspect ratios and wall-fluid slip lengths. Two immiscible fluids with identical properties are chosen in order to simplify the problem and focus only on how circulation is affected by slip at the wall.

The molecular dynamics simulations presented in this paper are performed using the LAMMPS package \citep{PlimptonS:95a}. The fluid's initial state is modelled as a face centered cubic (fcc) structure with the $x$ direction of the channel being aligned along the $[11\bar{2}]$ orientation of the face-centered cubic lattice. The wall is comprised of $2$--$3$ layers of atoms oriented along the $(111)$ plane of fcc lattice. The wall atoms are fixed to their lattice sites.

The pairwise interaction of molecules separated by a distance $r$ is modelled by the Lennard Jones (LJ) potential
\begin{equation}
V^{LJ}=4\epsilon\left[\left(\frac{\sigma}{r}\right)^{12}-\left(\frac{\sigma}{r}\right)^{6}\right],
\end{equation}
where $\epsilon$ and $\sigma$  are the characteristic energy and length scales. The potential is zero for $r>r_c$, and the cutoff radius, $r_c$ is $2.5 \sigma$.

The simulations are performed for immiscible fluids having properties same as that of liquid Argon. An interface between the two fluids is simulated by controlling the attractive force between the two. This is done by varying the LJ parameters. Reducing the value of $\epsilon$ reduces the attractive potential between the two fluids, while  increasing $\sigma$ allows to control the thickness of the interface. For the studies presented here the LJ parameters for fluid-fluid interaction are chosen as $0.2\epsilon$ and $3.0\sigma$.

\begin{table}
\begin{center}
\def~{\hphantom{0}}
\begin{tabular}{l*{4}{c}r} 
  & \AR & ~$H (\sigma)$~ & ~$L (\sigma)$~ & ~$W (\sigma)$~ \\[3pt]
  &1~ &~$52.9$~ &~$103.152$~ &~$144.4$~ \\
  &2~ &~$25.9$~ &~$103.152$~ &~$144.4$~ \\
  &3~ &~$25.9$~ &~$153.222$~ &~$144.4$~ \\
  \end{tabular}
  \caption{\label{tab:ar_h} Dimensions of the fluid domain in MD simulations, for droplets with different aspect ratio. $H$, $L$ and $W$ are height, length and width of the domain, respectively.}
\end{center}

\end{table}

The fluid is maintained in its equilibrium state having a number density $\rho=0.81\sigma^{-3}$ and temperature $T=1.1k_B/\epsilon$. The temperature is regulated by a thermostat which simulates the transfer of heat from the system to an external reservoir. A Langevin thermostat with a damping coefficient of $\Gamma=1.0\tau^{-1}$, where $\tau=\sqrt{m\sigma^{2}/\epsilon}$, is used here. The damping term is only applied to the $z$ direction to avoid biasing the flow. The equation of motion along the $z$ components is,
\begin{equation}
 m\ddot{z_i}=\sum_{j\neq i}\frac{\partial V{ij}}{\partial z_i}-m\Gamma \dot{z_i} + \eta_i.
\end{equation}
Here, $m$ is mass of the fluid molecule, $\sum_{j\neq i}$ denotes a sum over all interactions with $i$ and $\eta_i$ is a Gaussian distributed random force. The equations of motion were integrated using the Verlet algorithm \citep{VerletL:67a,TildesleyDJ:87a} with a time step $\Delta t=0.002\tau$.

The simulation is initially run for a time of $60\tau$ without any shear and then for an additional $\sim4000\tau$ with moving walls, allowing the flow to equilibrate. 
After $\sim4000\tau$, the spatial averaging is performed by dividing the fluid domain into square bins of size $\sim1.0 \sigma$ along the $x$--$y$ plane, and extending through the entire width of the channel. In addition to spatial averaging, ensemble averaging is done for a duration of $8000\tau$ in intervals of $200\tau$. In the case of non-wetting wall, averaging was done for an extended time of $16000\tau$. 


\section{Circulation model}\label{sec:circ_model}

A mathematical model that describes the effect of slip length on circulation, is established for a droplet moving in a channel where the two fluids are immiscible. The flow is driven by walls moving in opposite directions with a velocity, $U$, in the frame of reference of the droplet.

Velocity of fluid adjacent to the wall, $u^{wall}_f$, can be written from the Navier slip boundary condition
\begin{equation}
 u^{wall}_f=U-L_s \frac{d u_f}{dy},
\end{equation}
where $L_s$ is the slip length. In the case of a Couette flow, the $x$-component of velocity has a linear profile across the height of the droplet. Hence, the above equation can be re-written as
\begin{equation}
 u^{wall}_f = U \left(\frac{H}{H+2L_s}\right).
 \label{e:navier_slip}
\end{equation}

Circulation inside a droplet is expressed as
\begin{equation}
 \Gamma = \int_A^B \boldsymbol{u}_f \boldsymbol{\cdot} \mathrm{d}\boldsymbol{l} + \int_B^C \boldsymbol{u}_f \boldsymbol{\cdot} \mathrm{d}\boldsymbol{l} + \int_C^D \boldsymbol{u}_f \boldsymbol{\cdot} \mathrm{d}\boldsymbol{l} + \int_D^A \boldsymbol{u}_f \boldsymbol{\cdot} \mathrm{d}\boldsymbol{l},
\end{equation}
where A, B, C and D are the four corners of the droplet, see figure~\ref{fig:Schematic}.
Assuming the circulation contribution by the top and bottom walls and the left and right interfaces are the same, the equation reduces to
\begin{equation}
 \Gamma = 2 \int_0^{L} \boldsymbol{u}_f^{wall}\boldsymbol{\cdot} \mathrm{d}\boldsymbol{l} + 2 \int_0^{L_{intf}} \boldsymbol{u}_f^{intf}\boldsymbol{\cdot} \mathrm{d}\boldsymbol{l},
\end{equation}
where $L$ and $L_{intf}$ are the length of droplet along the wall and interface, respectively and ${u}_f^{intf}$ is the fluid velocity along the fluid-fluid interface. Assuming the velocity of fluid along the interface is same as that along the wall and substituting $u_f^{wall}$ onto the above equation, one obtains 
\begin{equation}
 \Gamma = 2 \int_0^{L} U \left(\frac{H}{H+2L_s(l)}\right) \mathrm{d}l + 2 \int_0^{L_{intf}} U \left(\frac{H}{H+2L_s(l)}\right) \mathrm{d}l.
\end{equation}

Performing a change of variable from $dl$ along the wall to $x$-direction and from $dl$ along the interface to $y$-direction we obtain,
\begin{equation}
 \Gamma = 2 \int_0^{L} U \left(\frac{H}{H+2L_s(x)}\right) \mathrm{d}x + 2 \int_0^{H} U \left(\frac{H}{H+2L_s(y)}\right) \frac{\mathrm{d}y}{\sin \theta},
\end{equation}
where $\theta$ is the respective averaged contact angle. Here, averaged contact angle is defined as the acute angle between the wall and a straight line which describes the fluid-fluid interface.

Scaling $x$ and $y$ with $L$ and $H$ respectively and doing a change of variable, circulation along a droplet can be represented as

\begin{equation}
 \frac{\Gamma}{2UL} = \int_0^{1}\frac{1}{1+2\frac{L_s(\zeta^*)}{H}} \mathrm{d}\zeta^* \left( 1+ \frac{1}{\AR\sin \theta}\right),
\end{equation}
where $\AR=L/H$ is the droplet aspect ratio.

The variation of $L_s$ can be assumed to be symmetric about the droplet centerline and can be approximated by a power law,
\begin{equation}
 \frac{\Gamma}{2UL} = 2 \int_0^{0.5}\frac{1}{1+2 \frac{L_s^o}{H}\left( \alpha \zeta^{*~\beta}+\gamma \right)} \mathrm{d}\zeta^* \left( 1+ \frac{1}{\AR\sin \theta}\right).
\end{equation}
Here, $\alpha, \beta$ and $\gamma$ are curve fitting coefficients to $L_s/L_s^o$ and $x$ data, where $L_s^o$ is the asymptotic value of slip length as defined by \citet{Troian:97a} for a single phase Couette flow.

Scaling the droplet circulation with slip at the walls, $\Gamma_s$, by the circulation with no-slip, $\Gamma_{ns}$, results in
\begin{equation}
 \frac{\Gamma_s}{\Gamma_{ns}} = 2 \int_0^{0.5}\frac{1}{\alpha' \zeta^{*~\beta}+\gamma'} \mathrm{d}\zeta^* 
 \left( \frac{1 + 1 /\AR\sin \theta_s}{1+ 1 /\AR\sin \theta_{ns}}\right),
 \label{e:model}
\end{equation}
where $\alpha'=2 \frac{L_s^o}{H}\alpha$, $\gamma'=2 \frac{L_s^o}{H}\gamma+1$ and $\theta_s$, $\theta_{ns}$ are the average contact angles for wall with slip and no-slip, respectively.

For a long droplet, the decrease in fluid velocity in the vicinity of the triple contact point has a negligible affect on the total droplet circulation. Hence, slip length can be assumed to be a constant along the length of the droplet. In this case the above equation reduces to,
\begin{equation}
 \frac{\Gamma_s}{\Gamma_{ns}} = \frac{1}{1+2 \frac{L_s^o}{H} } \left( \frac{1 + 1 /\AR\sin \theta_s}{1+ 1 /\AR\sin \theta_{ns}}\right).
\end{equation}

From the above expression it can be seen that circulation varies inversely with slip length, though the extent of its effect on circulation is dependent on whether the length scale of the problem is comparable to the slip length scale. Also, it can be observed that for droplets with high \AR~, the contribution to circulation from the fluid-fluid interface reduces relative to the contribution from the wall.

\section{Results}\label{sec:results}

Several runs are performed for five different cases of wall-fluid properties, as described in table~\ref{tab:myfirsttable}, with decreasing wall wettability from case 1 to case 5.  The walls move with a speed of $U= 0.1\sigma/\tau$ in opposite directions and the velocities from MD presented in this paper are normalised by it. In order to compute various values at the wall, a reference plane is defined at a distance of $0.5\sigma_{wf}$ away from the wall lattice site.

The continuum simulations simulate a rectangular droplet with $90\degree$ contact angle and a free surface boundary condition at the interface. The top and bottom walls move in opposite directions. The length of the droplet is fixed at $L=1$ and the height is calculated based on the aspect ratio. Details of the Navier Stokes continuum solver is given in \citet{Mohseni:12n}. A set of runs are made with varying slip lengths and \AR.

\begin{figure}
\begin{center}
\includegraphics[width=0.53\textwidth]{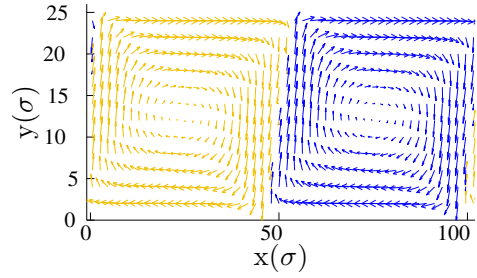}
\end{center}
\caption{MD result for the velocity field of a droplet moving in a channel. Each droplet exhibits a co-rotating circulatory flow. The y-axis is stretched to provide clarity.}
\label{f:vel_field} 
\end{figure}

  The fluid-fluid interface in the droplet forces convection normal to the wall, causing the fluid to circulate inside a moving droplet. A plot of the velocity field of a moving droplet in figure~\ref{f:vel_field} shows the presence of circulation inside the droplet. As the two fluids share the same wall they have a co-rotating flow. The flow field is similar to what one would see in continuum scale droplets \citep{PamulaVK:03b,Mohseni:08g} and was previously observed by \citet{KoplikJ:89a} and \citet{ThompsonP:89a} in MD simulations.

\begin{table}
\begin{center}
\def~{\hphantom{0}}
\begin{tabular}{l*{4}{c}r} 
  & Case & ~$\epsilon^{wf}/\epsilon$~ & ~$\sigma^{wf}/\sigma$~ & ~$\rho_w/\rho$~ \\[3pt]
  &1~ &~$1.0$~         & ~$1.0$~ & ~$1$~ \\
  &2~ &~$0.6$~         & ~$1.0$~ & ~$1$~ \\
  &3~ &~$0.6$~         & ~$0.75$~ & ~$4$~ \\
  &4~ &~$0.4$~         & ~$0.75$~ & ~$4$~ \\
  &5~ &~$0.2$~         & ~$0.75$~ & ~$4$~ \\
  \end{tabular}
  \caption{\label{tab:myfirsttable} Five different cases of wall-fluid properties are considered, with wall wettability decreasing from case 1 to 5.}
\end{center}

\end{table}

\begin{figure}
\begin{center}
  \includegraphics[width=0.43\textwidth]{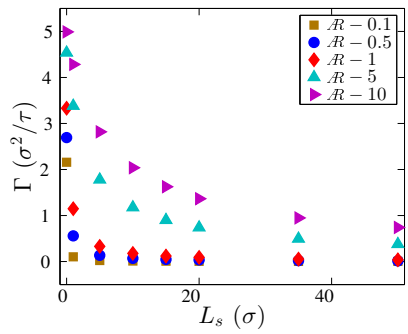}
\end{center}  
  \caption{Circulation versus slip length for droplets with different \AR. Results from continuum simulations shows the decrease in circulation with increasing slip length which is a measure of wettability.}
  \label{f:Circ_ls}
\end{figure}

 For the Couette flow problem considered here, wall movement drives the circulation inside a droplet, hence it is intuitive that the occurrence of slip at the wall will affect circulation. Partial slip at the wall reduces circulation and if there is perfect slip the circulation reduces to zero. This reduction in circulation directly impacts the rate of mixing, chemical reaction, mass and heat transfer essential for various droplet based microfluidic devices. Figure~\ref{f:Circ_ls} shows how total droplet circulation varies inversely with slip length, which could be a measure of  wall wettability. Its variation with slip length is considered for droplets with different aspect ratios.

\begin{figure}
\centering
\begin{minipage}{0.04\linewidth}\begin{center}  (a) \end{center}  \end{minipage}
\begin{minipage}{0.4\linewidth}\begin{center}
 \includegraphics[width=0.9\linewidth]{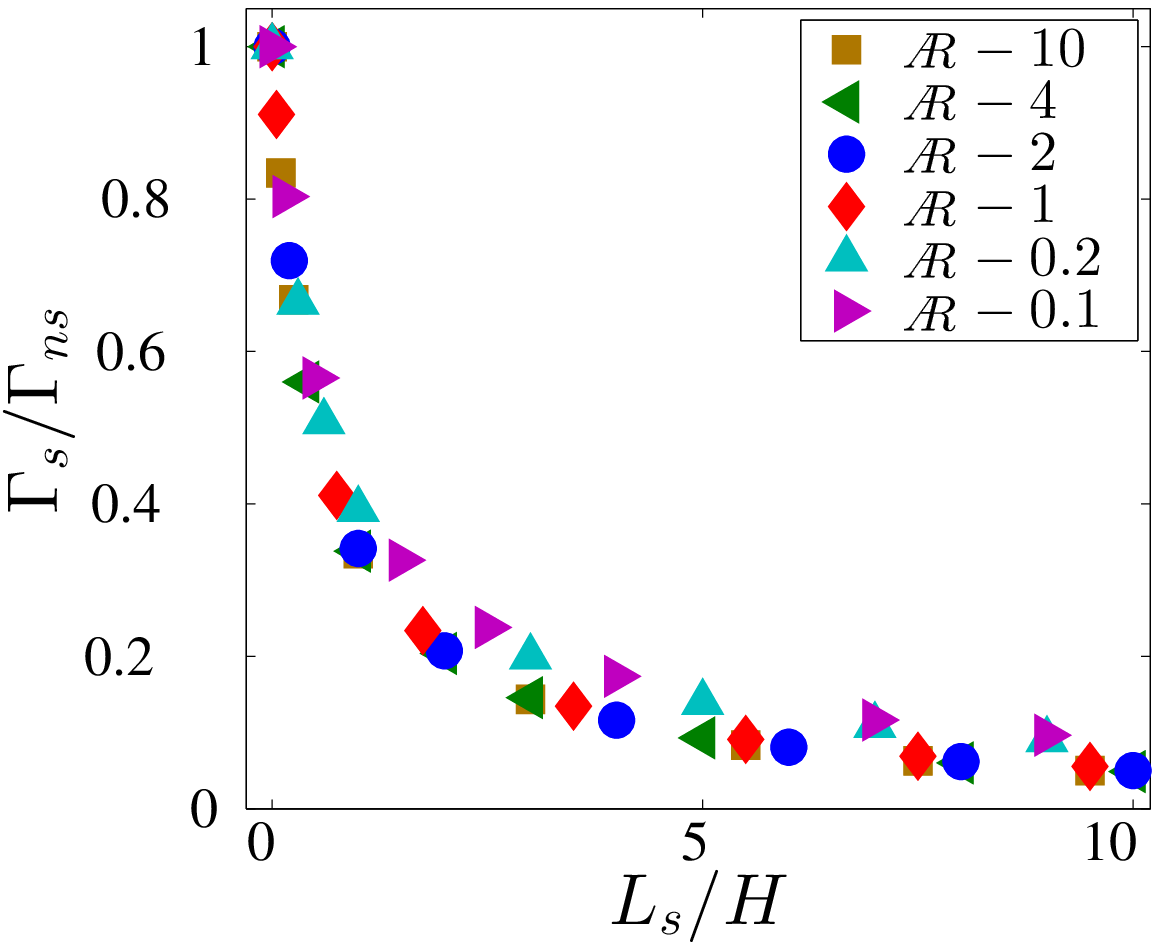}
\end{center}\end{minipage}
\begin{minipage}{0.04\linewidth}\begin{center} (b) \end{center}\end{minipage} 
\begin{minipage}{0.4\linewidth}\begin{center}
 \includegraphics[width=0.9\linewidth]{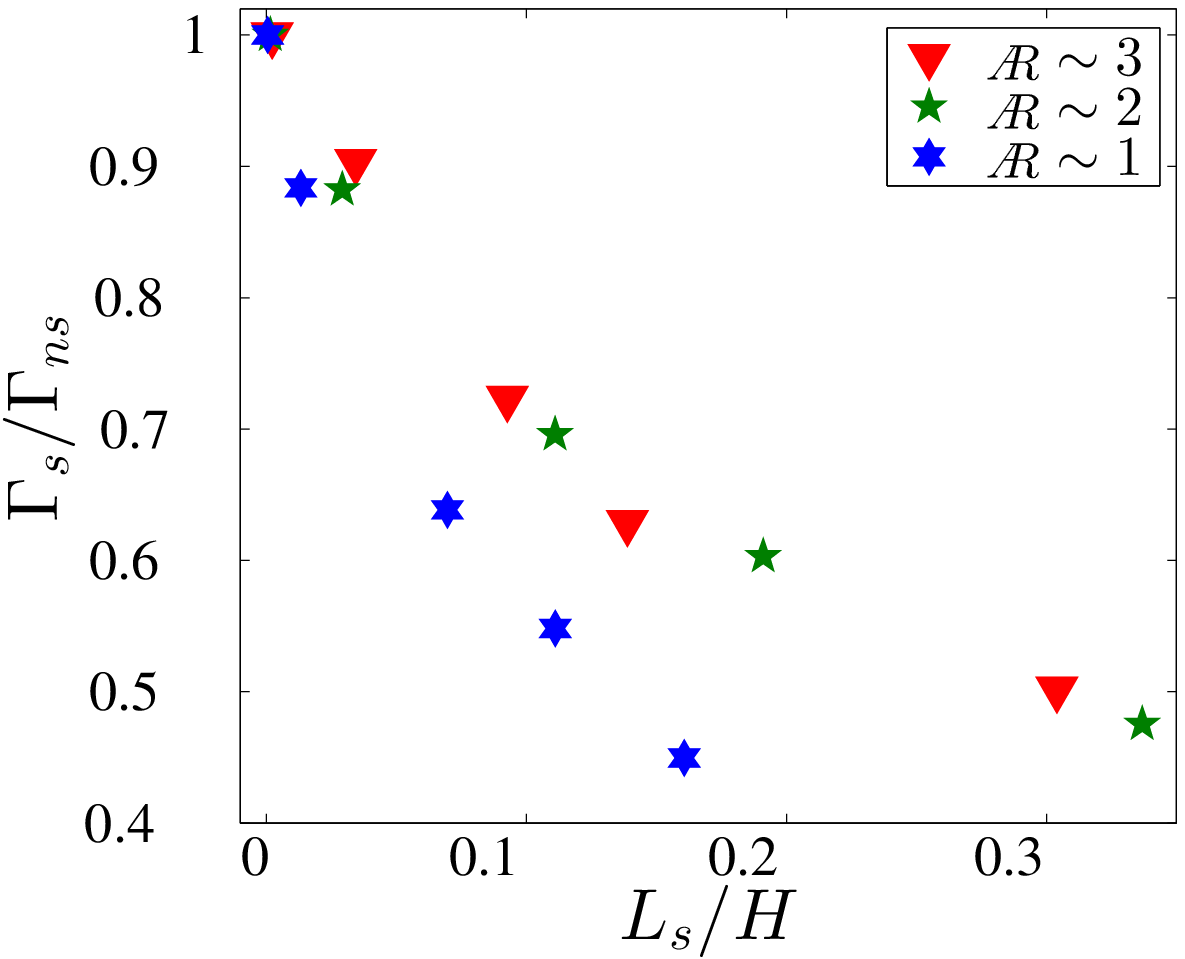}
\end{center}\end{minipage} 

\caption{\footnotesize Nondimensional circulation versus slip length for, (a) continuum simulations and (b) MD simulations.
Results from continuum and MD simulations for various droplet aspect ratios are shown. For droplet with high \AR~ the results collapse onto a curve, confirming the proper scaling of circulation and slip length. For low \AR~ droplets the data does not collapse onto the same curve as that for high \AR~ droplets.}
  \label{f:Circ_ls_collapse}
\end{figure}

\begin{figure}
\begin{center}
  \includegraphics[width=0.43\textwidth]{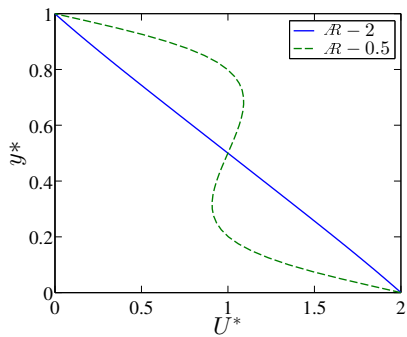}
\end{center}  
  \caption{Velocity profile across the center of the droplet for \AR~ of 0.5 and 2.0. Here, $y^*=y/H$ and $U^*=|u/U-1|$, which gives the velocity in the lab frame of reference. Deviation from the linear velocity profile of a Couette flow is seen which also implies the formation of two circulatory flows in a droplet.}
  \label{f:Vel_comp}
\end{figure}

The proper scaling for circulation and slip length is identified from the circulation model. Circulation with slip, $\Gamma_s$, is scaled by circulation for a perfectly wetting (no-slip) wall, $\Gamma_{ns}$, and slip length, $L_S$, is scaled by the droplet height, $H$. For droplets with high \AR~, the data for circulation collapse onto a single curve, see figure~\ref{f:Circ_ls_collapse}. From the MD results presented in  figure~\ref{f:Circ_ls_collapse}(b), a decrease in circulation of as much as $50\%$ is seen for a non-wetting case, demonstrating the extent of effect that slip can have on circulation. The impact of length scale of the problem  on circulation can also be seen here. For droplets where the height of the droplet is comparable to slip length, there is a larger reduction in circulation in comparison to a taller droplet. For low \AR~ droplets, $\AR\leq0.2$ for continuum results and $\AR\leq1$ for MD results, the data deviates from the collapsed 
curve. This deviation can be attributed to the change in flow dynamics of tall droplets. As figure~\ref{f:Vel_comp} shows, for tall or narrow droplet a pair of circulatory flows form inside the droplet. This breaks down the assumption of a linear velocity profile across the height of the droplet.

\begin{figure}
\begin{center}
  \includegraphics[width=0.43\textwidth]{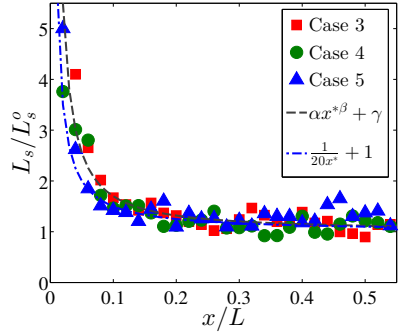}
\end{center}  
  \caption{Non-dimensional slip length along the wall. Plot shows the collapse of data for various cases of wall wettability. A power law of the form $\alpha x^{*\beta} +\gamma$, where $x^*=x/L$, is a good fit for this data.}
  \label{f:ls_collapse}
\end{figure}

The variation of slip length along the wall needs to be described in order to calculate circulation based on the model established in the previous section. It is found that by scaling slip length, $L_s$, by its asymptotic value, $L_s^o$, the data for different cases of wall wettability collapse onto a single curve, see  figure~\ref{f:ls_collapse}. A power law of the form $\alpha x^{*\beta} +\gamma$, where $x^*=x/L$, is a good fit for this data. The distribution of slip length can be assumed to be symmetric about the center of the droplet. The coefficients of the power law are obtained by nonlinear least squares method as, $\alpha=0.0468$, $\beta=-1.166$ and $\gamma=1.013$. They can be rounded off to $\alpha=0.05$, $\beta=-1.0$ and $\gamma=1.0$, to obtain ${L_s}/{L_s^o}=({1}/{20}){L}/{x}+1$.

\begin{figure}
\centering
\begin{minipage}{0.04\linewidth}\begin{center}  (a) \end{center}  \end{minipage}
\begin{minipage}{0.4\linewidth}\begin{center}
 \includegraphics[width=0.9\linewidth]{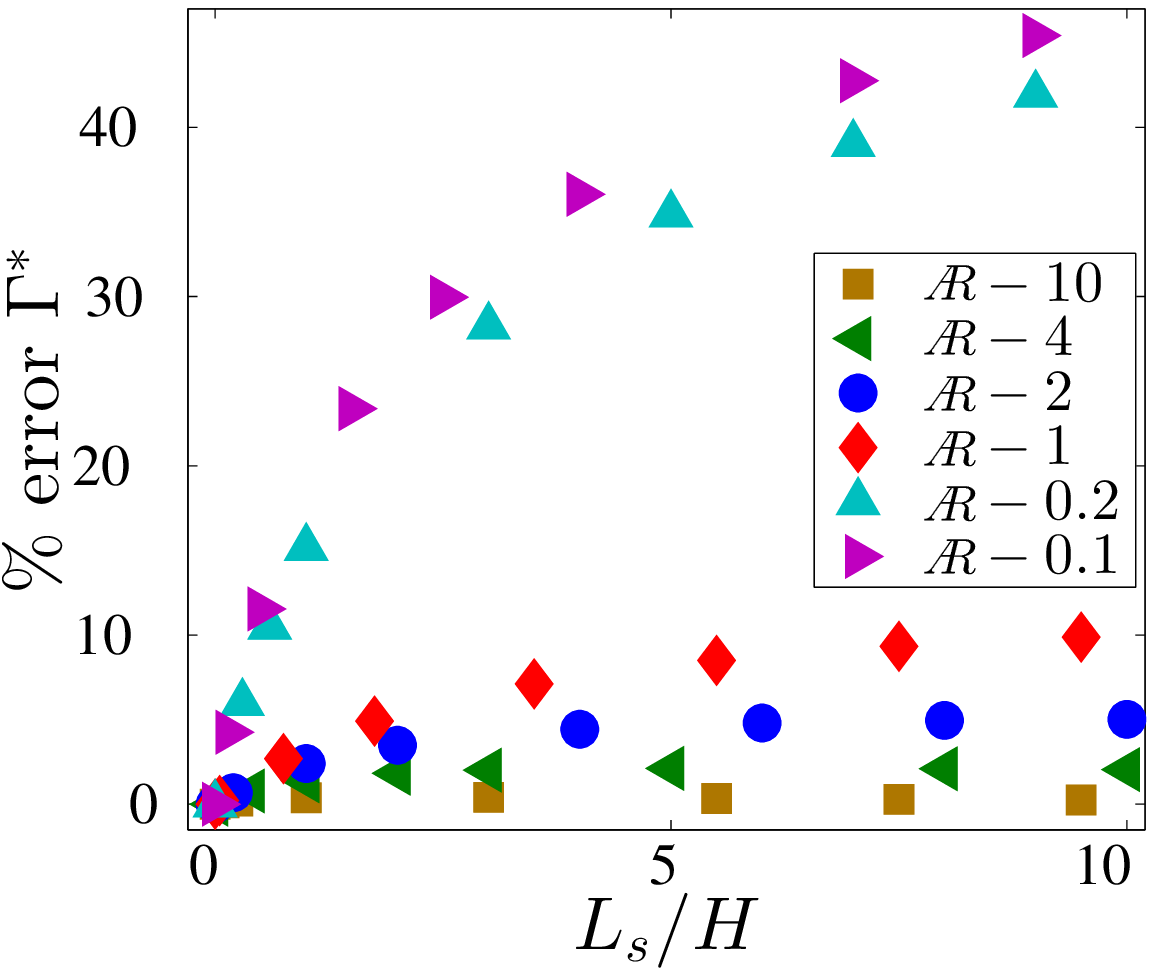}
\end{center}\end{minipage}
\begin{minipage}{0.04\linewidth}\begin{center} (b) \end{center}\end{minipage} 
\begin{minipage}{0.4\linewidth}\begin{center}
 \includegraphics[width=0.9\linewidth]{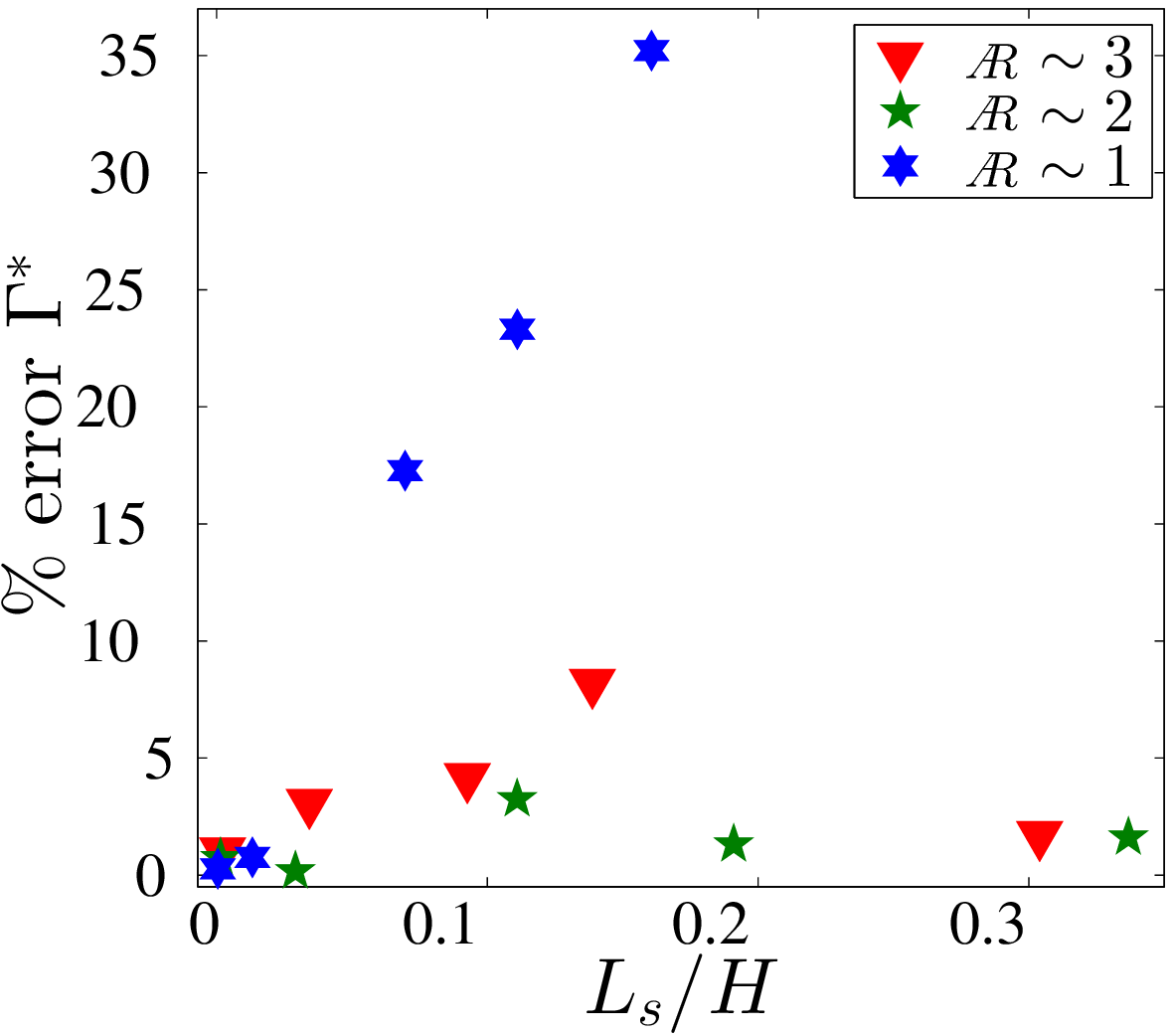}
\end{center}\end{minipage} 

\caption{\footnotesize $\%$ error of non-dimensional circulation in (a) continuum simulations and (b) MD simulations.
The percentage relative error for circulation calculated using the circulation model is plotted for various droplet \AR. The error is computed as, $(|\Gamma^*_{\text{model}}-\Gamma^*_{\text{simulation}}|)/\Gamma^*_{\text{simulation}}$, where $\Gamma^*=\Gamma_s/\Gamma_{ns}$. The error increases with decrease in droplet \AR~ and increase in slip length. For large aspect ratio droplets, the error is less than $10\%$.}
  \label{f:error}
\end{figure}

The circulation model established in the previous section is verified by comparing the scaled circulation ($\Gamma_s/\Gamma_{ns}$) obtained from MD and continuum simulations with that calculated from the model. Figure~\ref{f:error} shows the model to be accurate for droplets with high \AR, having an error of less than $10\%$. The accuracy of the circulation model decreases sharply for droplets with low \AR. This is mostly attributed to the breakdown of assumption that the velocity profile is linear across the height of the droplet, as discussed previously. In addition, the error also increases gradually with slip length until it reaches an asymptotic value. The rate of increase in error is seen to increase with decrease in aspect ratio. This is a result of the assumption made that fluid velocity at the interface is the same as that at the wall. But, in fact the fluid moving along the fluid-fluid interface decays due to viscous diffusion as 
it travels towards the center of the droplet. Hence, decrease in fluid inertia due to decrease in wettability and/or travelling longer distances along the fluid-fluid interface in the case of droplets with lower \AR~ leads to reduction in accuracy of the model. From the circulation model given by equation \ref{e:model} it is seen that the contribution of velocity along the interface to total circulation increases with decreasing aspect ratio. Although, there is a reduction in accuracy of the model for low \AR~ droplets, it is known that for most practical applications, high \AR~ droplets are used for which the model is accurate.

In the current study only the case of immiscible fluids were considered but slip at a general fluid-fluid interface of two partially miscible fluid could also have an affect on circulation in a droplet. Immiscible fluids have a sharp, distinct interface where the two fluids do not experience any transfer of tangential momentum at the interface. This is analogous to a fluid with perfect slip at the wall. In the case of partially miscible fluids the interaction between the two fluids leads to a change in momentum along the interface. As the flow in the droplets are co-rotating there could be an apparent reduction in total circulation in a droplet, depending on the relative molecular mass of the two fluids. But, as circulation is calculated over a material line, the occurrence of mass transfer across the interface leads to difficulties in its calculation. Using MD, it is currently computationally expensive to obtain sufficient statistics for a case with partial miscibility between the fluids as the interface starts stretching, eventually leading to breakup of the droplet.

\section{Conclusion}
Circulation is an important factor that affects mixing, chemical reaction and heat transfer inside a droplet, which have many practical applications. This paper demonstrates the presence of circulation in droplets using MD simulations and its inverse dependence on slip length at the wall-fluid interface. A mathematical model for circulation is established that shows this dependence on slip length,
$$\frac{\Gamma_s}{\Gamma_{ns}} = 2 \int_0^{0.5}\frac{1}{2\frac{L_s}{H}\alpha \zeta^{*~\beta}+\gamma'} \mathrm{d}\zeta^* 
 \left( \frac{1 + 1 /\AR \sin \theta_s}{1+ 1/\AR \sin \theta_{ns}}\right).$$
The model also shows that the extent of change in circulation caused by slip depends on whether the length scale of the problem is of the order of slip length scale. On scaling circulation and slip length, the results for different droplet \AR~ collapse onto a single curve. The model is verified using results from continuum and MD simulations and is shown to be accurate for droplets with high \AR. Also, it is seen that there can be as much as $50\%$ change in circulation for a droplet moving in a non-wetting channel.\\

This research was supported by the Office of Naval Research.


\bibliographystyle{jfm}

\bibliography{RefA1}

\end{document}